\documentclass[aps,pre,twocolumn,showpacs,showkey,square,numbers,amssymb,amsmath,longbibliography]{revtex4-1}
\usepackage{bm}
\usepackage{times,float}
\usepackage{graphicx}
\usepackage[usenames,dvipsnames,svgnames]{xcolor}
\usepackage{hyperref}
\hypersetup{colorlinks=true, linkcolor=NavyBlue, citecolor=PineGreen,urlcolor=cyan}

\begin{document}	
\title{The spectral density of dense random networks  and the breakdown of the Wigner semicircle law}
\author{Fernando L. Metz}
\email[]{fmetzfmetz@gmail.com}
\address{Physics Institute, Federal University of Rio Grande do Sul, 91501-970 Porto Alegre, Brazil}
\address{London Mathematical Laboratory, 18 Margravine Gardens, London W6 8RH, United Kingdom}

\author{Jeferson D. Silva}
\address{Physics Institute, Federal University of Rio Grande do Sul, 91501-970 Porto Alegre, Brazil}

\begin{abstract}
  Although the spectra of random networks have been studied for a long time, the influence of network topology on the dense
  limit of network spectra remains poorly understood.  
  By considering the configuration model of networks with four distinct  degree distributions, we show that the spectral density of the adjacency
  matrices of dense random networks  is determined by the strength of the degree fluctuations.
  In particular, the eigenvalue distribution of dense networks with an exponential degree distribution
  is governed by a simple equation, from
  which we uncover a logarithmic singularity in the spectral density. We also derive a relation between the fourth moment of the eigenvalue
  distribution and the variance of the degree distribution, which leads to a sufficient condition for
  the breakdown of the Wigner semicircle law for dense random networks. Based on the same relation, we propose a classification scheme of the distinct universal behaviours of the spectral density
  in the dense limit. Our theoretical findings should lead to important insights on the mean-field behaviour of models
  defined on graphs.
\end{abstract}

\maketitle

\section{Introduction}
 
A random network or graph is a collection of nodes joined by edges following a probabilistic rule. Random
networks are formidable tools to model  large
assemblies of interacting units, like neurons in the brain, computers and routers in the Internet, or
persons forming a friendship network \cite{NewmanBook}. Motivated by our increasing ability to
collect and process vast amounts of empirical data,  the theory of random networks
has experienced an enormous progress, leading
to important insights in physics, biology, and sociology \cite{bornholdt2003}.
The implications of the structure of networks to the dynamical processes occurring on
them remains a fundamental topic in network theory \cite{BarratBook}.

Dynamical processes on a random network are to a large extent governed by the  spectrum
of the corresponding adjacency random matrix. This is a random matrix where each entry equals the strength of the interaction
between a pair of nodes.
The study of a broad range of problems amounts to linearize a large set of differential equations \cite{Metz2019,Neri2019,Tarnowski2020}, coupled through a random network, around a
stationary state, whose stability and transient dynamics are ultimately determined by the eigenvalue distribution of the
adjacency matrix. Examples in this context are the study of the epidemic threshold for
the spreading of diseases \cite{Pastor2015}, the synchronization transition in networks of coupled oscillators \cite{Restrepo2005,Arenas2008}, and the functional stability
of large biological systems, such as gene regulatory
networks \cite{Chen2019,Guo2020}, ecosystems \cite{May,Allesina2015,Grilli2016}, and neural networks\cite{Sompolinsky1988,Rajan2006, Schuessler2020}.
The spectrum of the adjacency matrix also contains information about the network
structure, since the trace of the adjacency matrix raised to a certain power yields the number of network loops of a given length \cite{VanBook}.
Therefore, the problem of how the network architecture influences the spectrum of the adjacency matrix has attracted considerable
research efforts \cite{Farkas2001,Doro2003,Rodgers2005,Rogers2010,Kuhn2011,Metz2011,Metz2019,Newman2019}.

Synthetic models of random networks provide a controlled  way to study the role
of the network structure on the spectrum of the adjacency matrix.
The degree sequence is the most basic tool to characterize the graph structure \cite{NewmanBook}. The degree of a node
counts the number of edges attached to the node, while the degree sequence specifies
all network degrees. When a network has an infinitely large number of constituents, it is natural
to consider the degree distribution, i.e. the fraction of nodes with a certain degree, instead of dealing with the
degree sequence. The configuration model stands out as one of the most fundamental and versatile models
of random graphs \cite{Molloy95,Molloy98,NewmanBook,Newman2001}, since it enables the degree distribution to be freely specified, while keeping the
pattern of interconnections entirely random.
From a practical viewpoint, the configuration model resolves one
of the main shortcomings of Erd\"os- R\'enyi random graphs \cite{Solomon51,Erdos59,Erdos60}, namely its Poisson degree distribution, which has
little resemblance with the long-tailed degree distributions found in empirical networks \cite{Albert2002,Clauset2009}.
The priceless possibility to fix the degree
distribution of a random graph is not only useful to model the structure of empirical networks, but it offers
the ideal setting to examine the impact of degree fluctuations on the spectrum of the adjacency matrix.

There has been a significant amount of numeric and analytic work on the spectral properties of adjacency random matrices
\cite{Rodgers1988,Farkas2001,Goh2001,Doro2003,Rodgers2005,Rogers2008,Kuhn2008,Rogers2009,Rogers2010,Metz2011,Kuhn2011,Neri2012,Metz2016,Metz2019,Newman2019,Newman2019A}. A
remarkable analytic result is the set of exact and mathematically rigorous equations determining the
eigenvalue distribution of the configuration model \cite{Rogers2008,Kuhn2008,Rogers2009,Bordenave2010}.
These equations form a natural starting point to study the impact of degree fluctuations on network spectra. 
Unfortunately, aside from a few particular cases \cite{Bordenave2010,Neri2012,Metz2011}, these equations can be analytically solved
only in the dense limit \cite{Rogers2008,Kuhn2008}, when the mean degree becomes infinitely large and random networks
approach a fully-connected structure.

The analytic solution for the dense limit  of the aforementioned equations is at the root of perturbative and non-perturbative approximations for
the eigenvalue distribution of large-degree random networks \cite{Rodgers1988,Semerjian2002,Doro2003}.
With the exception of graphs with a power-law degree distribution \cite{Farkas2001,Goh2001,Rodgers2005}, whose moments are divergent, one expects that the
eigenvalue distribution of undirected random networks converges, in the dense limit, to the Wigner semicircle distribution of random
matrix theory, reflecting a high level of universality. This expectation has been 
confirmed numerically \cite{Farkas2001} and analytically \cite{Rogers2008,Kuhn2008} for Erd\"os-R\'enyi and regular random
graphs. However, the recent work \cite{Dembo} has rigorously shown that the spectrum of the
  configuration model does depend on the degree distribution in the dense limit. The main theorem in \cite{Dembo}  also provides an approach to
  compute the spectral density for specific degree distributions. 
In spite of these rigorous results, a more detailed understanding of how the structure of a dense random network
influences its spectrum, and the universal status of the Wigner semicircle law, is still lacking.

In this work we study the dense limit of the eigenvalue distribution of undirected networks drawn from the configuration model. We analyze 
four examples of degree distributions in which all moments are finite and the variances scale differently with the average degree, and
we show that the dense limit of the
eigenvalue distribution is determined by the degree fluctuations. 
In particular, we derive an equation yielding the eigenvalue distribution of dense random graphs
with an exponential degree distribution \cite{Newman2001}, from which we unveil the existence of a logarithmic divergence in the spectral density and the absence
of sharp spectral edges, i.e., the eigenvalue distribution of exponential random graphs is supported on the entire real line.
These are remarkable differences with respect to the Wigner semicircle distribution of random matrix theory \cite{VivoBook}. We also discuss how
the analytic equations for the spectral density can be derived in two different ways: from the dense limit of the exact resolvent equations
in \cite{Rogers2008,Kuhn2008,Bordenave2010}, and from the main theorem proved in \cite{Dembo}.
Based on an exact calculation of the fourth moment of the eigenvalue distribution, we obtain a sufficient condition
for the breakdown of the Wigner semicircle law. This condition is given only in terms of the variance of the degree distribution. We
finish the paper by proposing a classification of the different universal behaviours of the eigenvalue
distribution of the configuration model in the dense limit, based on an exponent characterizing  how the variance of the degree
distribution diverges for increasing mean degree. 

Our paper is organized as follows. In the next section we define the adjacency matrix of random networks and
the spectral density. Section \ref{confmodel} introduces the configuration model and defines the
degree distributions studied in this work. In section \ref{bbba}, we present numeric and
analytic results for the dense limit of the spectral density for each example of degree distribution. Section
\ref{bbba} explains how the analytic equations for the spectral density are obtained from the dense limit of the
resolvent equations, while in section \ref{rigor} we derive these equations from the main theorem in reference \cite{Dembo}.
The condition for the breakdown of the Wigner semicircle law and the classification of the different universal behaviours
are discussed in section \ref{assa}. We summarize our results and conclude in section \ref{gugu}.

\section{The adjacency matrix of random networks}

Let $N$ be the total number of network nodes. The set
of binary random variables $\{ C_{ij} \}_{i,j=1,\dots,N}$ specifies the network structure: if $C_{ij} =1$, there
is an undirected link  $i \leftrightarrow j$ between nodes $i$ and $j$, while $C_{ij} =0$ means this link is absent.
We also associate a random weight
$J_{ij} \in \mathbb{R}$ to each edge $i \leftrightarrow j$, which accounts for the interaction strength between nodes $i$ and $j$. 
We consider undirected simple random networks that have no self-edges nor multi-edges, in which $C_{ij}= C_{ji}$, $J_{ij}=J_{ji}$, and $C_{ii} = 0$. 

The degree $K_i = \sum_{j=1}^N C_{ij}$  of node $i$ is a random variable
that counts the number of nodes attached  to $i$, while the degree sequence  $K_1,\dots,K_N$ provides global information
on the fluctuations of the network connectivity.
In the limit $N \rightarrow \infty$, it is more
convenient to work with the degree distribution
\begin{equation}
p_k = \lim_{N \rightarrow \infty} \frac{1}{N} \sum_{i=1}^N \delta_{k,K_i} ,
\end{equation}  
where $\delta$ is the Kronecker symbol.  The average degree $c$ and the variance $\sigma^2$ of $p_k$ read
\begin{equation}
c = \sum_{k=0}^\infty k p_k, \qquad
\sigma^2 = \sum_{k=0}^\infty p_k \left( k - c  \right)^2. \nonumber
\end{equation}  
The degree distribution $p_k$ is one of the primary quantities to characterize the structure
of random networks in the limit $N \rightarrow \infty$.

We will study the eigenvalue distribution of the $N \times N$ symmetric adjacency matrix $\boldsymbol{A}$, with elements
defined as 
\begin{equation}
  A_{ij} = \frac{C_{ij} J_{ij}}{\sqrt{c}}.
  \label{gaga}
\end{equation}  
The adjacency matrix fully encodes the network structure, along with the coupling strengths between adjacent nodes.  
The empirical spectral measure of $\boldsymbol{A}$ is given by
\begin{equation}
  \rho_N( \lambda) =   \frac{1}{N} \sum_{\alpha=1}^N \delta \left( \lambda - \lambda_{\alpha} \right),
  \label{jkla}
\end{equation}  
where $\lambda_1,\dots,\lambda_N$ are the (real) eigenvalues of $\boldsymbol{A}$.
By introducing the $N \times N$ resolvent matrix
\begin{equation}
\boldsymbol{G}(z) = \left( z  - \boldsymbol{A} \right)^{-1},
\end{equation}  
with $z = \lambda - i \eta$ on the lower half complex plane, the empirical spectral measure follows from the
diagonal elements of $\boldsymbol{G}(z)$
\begin{equation}
  \rho_N( \lambda) = \frac{1}{\pi N} \lim_{\eta \rightarrow 0^{+}}  \sum_{i=1}^N  {\rm Im} G_{ii} (z) .
  \label{popo1}
\end{equation}
In the limit $N \rightarrow \infty$, the empirical mean of ${\rm Im} G_{ii}(z)$ normally converges to
its ensemble averaged value, obtained from the distribution of $\boldsymbol{A}$, which implies
that $\rho(\lambda) = \lim_{N \rightarrow \infty} \rho_N( \lambda)$ is well-defined.
Here we will study the
spectral density $\rho(\lambda)$ when the average degree grows to infinity, hence the limit $c \rightarrow \infty$
is performed {\it after} $N \rightarrow \infty$. The scaling of the elements $A_{ij}$ with the average degree $c$ in Eq.~(\ref{gaga})
ensures the spectral density has a finite variance when $c \rightarrow \infty$. 

\section{The configuration model of networks} \label{confmodel}

We study random networks with arbitrary degree
distributions, known as the configuration model of networks \cite{Molloy95, Molloy98,Newman2001,NewmanBook}, where $p_k$ is specified
at the outset.
A single instance of the adjacency matrix of the configuration model is created as follows. First, the degrees $K_1,\dots,K_N$ are drawn independently
from a common distribution $p_k$. After assigning $K_i$ stubs of edges to each node $i$ ($i=1,\dots,N$), a pair of stubs is uniformly chosen at
random and then connected to form an edge.
This last step is repeated on the remainder stubs until there are no stubs left, and the outcome is a particular
matching of the stubs with the prescribed random degrees $K_1,\dots,K_N$.
We do not allow for the existence of self-edges and multi-edges in the random network.
We set $A_{ij} = J_{ij}/\sqrt{c}$ if there is an edge connecting nodes $i$ and $j$, and $A_{ij}=0$ otherwise.
The coupling strengths $\{ J_{ij} \}_{i,j=1,\dots,N}$ are i.i.d. random variables 
drawn from a common distribution $P_J$.
We refer to \cite{NewmanBook} for other properties and subtleties of the configuration model.
In the limit $N \rightarrow \infty$, the ensemble of adjacency random matrices $\boldsymbol{A}$ constructed from this procedure is
specified by the degree distribution $p_k$ and the distribution of weights $P_J$.
This is probably the simplest network model that allows to clearly exploit the influence of degree fluctuations
on the spectral properties of $\boldsymbol{A}$.

We will present results for regular, Poisson, exponential, and Borel degree distributions. The analytic expression
for $p_k$ in each case, together with the variance $\sigma^2$, is displayed in table \ref{univ}. The
properties of the configuration model with Poisson and exponential degree distributions have been extensively discussed in \cite{Newman2001,NewmanBook}. The
degree distributions in table \ref{univ} obey the following properties: (i) $p_k$ is parametrized solely in terms of the mean degree $c$; (ii) all moments of $p_k$ are finite
for $c < \infty$; (iii) for sufficiently large $c$, the variances obey $\sigma_{\rm reg}^2 < \sigma_{\rm pois}^2 < \sigma_{\rm exp}^2 < \sigma_{\rm bor}^2$.
These four examples of degree distributions are very convenient to study, in a controllable way, the effect of degree fluctuations on
random networks as we increase $c$. Indeed,  we can compare for $c \gg 1$ the spectral density of networks with the same average
degree, but with increasing variances $\sigma^2$. We do not consider here random networks with power-law degree
distributions \cite{Albert2002,Clauset2009}, since in this case higher-order moments of $p_k$ diverge already for finite $c$.
\begin{table}[ht]
\centering 
\begin{tabular}{c c c c c}
\hline\hline 
 & Regular & Poisson & Exponential & Borel \\ [0.5ex] 
\hline 
$p_k$  & $\delta_{k,c}$  &  $\frac{e^{-c} c^k}{k!}$   &  $\frac{1}{c+1} \left( \frac{c}{c+1} \right)^k $
& $\frac{e^{-\frac{(c-1)}{c}k}}{k! }   \left[ \frac{k (c-1)}{c} \right]^{k-1}   $  \\ 
$\sigma^2$ & $0$ & $c$ & $c^2 + c$ & $c^3 - c^2$ \\
[1ex] 
\hline 
\end{tabular}
\caption{The analytic expression for the degree distribution $p_k$ and the corresponding variance $\sigma^2$ in the case of regular,
  Poisson, exponential, and Borel random networks. The Borel degree distribution is defined for $c > 1$, while the
  other three distributions are defined for $c >0$. The Borel degree distribution is supported in $k \geq 1$, the degree distribution
  of regular graphs is supported at $k=c$, and both Poisson and
exponential degree distributions are supported in $k \geq 0$.}
\label{univ} 
\end{table}

\begin{figure}[h]
  \begin{center}
    \includegraphics[scale=1.1]{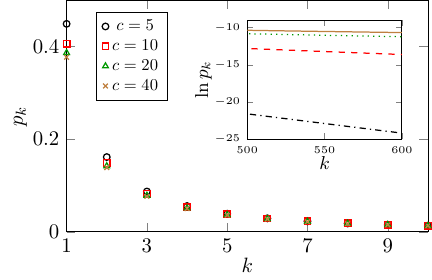}
    \caption{The Borel degree distribution $p_k$ (see table \ref{univ}) for different values of the average degree $c$. The inset
      depicts the exponential tail of $p_k$ for large $k$. The values of $c$ in the inset are $c=40$ (solid line), $c=20$ (dotted line), $c=10$ (dashed line), and
      $c=5$ (dot-dashed line).
}
\label{boreld}
\end{center}
\end{figure}

Although the Borel degree distribution is not commonly employed in the
study of random networks, the rate of its exponential decay is smaller than the one of the exponential degree distribution with the same $c$, which makes the Borel
distribution well-suited to our analysis.
The Borel distribution, introduced
in the context of queuing theory \cite{Borel1942,ConsulBook}, also appears as
the distribution of the total progeny in a Galton-Watson branching process with Poisson 
distributed degrees \cite{Finner2015}.
Figure \ref{boreld} illustrates the Borel degree distribution
for different average degrees. For fixed $k \ll c$, the Borel distribution $p_k$ attains a finite limit as
$c \rightarrow \infty$, in contrast to the Poisson and  the exponential degree distribution. For $k = O(c)$, the
Borel distribution $p_k$ is proportional to $1/\sqrt{c}$ for $c \gg 1$. Therefore, Borel networks with
large $c$ contain a finite fraction of nodes with $c$-independent degrees and a smaller fraction of nodes
with degrees proportional to $c$. Note also that $p_k$ in figure \ref{boreld} is highly skewed, its mode is located at $k=1$, and the
tail of $p_k$ decays as $\ln p_k \sim \left( \frac{1}{c} + \ln\left(  \frac{c-1}{c} \right)   \right) k$ for $k \gg 1$.
All network models considered here have exponentially decaying degree
distributions.

\section{The dense limit of the spectral density } \label{bbba}

The configuration model of networks has a key property:  in the limit $N \rightarrow \infty$, the set of nodes in the
neighborhood of a given node, drawn uniformly from the network, is arranged in a tree-like structure \cite{Bordenave2010}.
Nevertheless, the topology of random networks is fundamentally different from a Cayley tree \cite{BaxterBook}, in the sense that
boundary nodes are absent from the configuration model and cycles do survive in the limit $N \rightarrow \infty$, but their average length scales
as $\ln N$ for $N \gg 1$. Roughly speaking, a random network can be seen as a Cayley tree with fluctuating degrees, wrapped onto itself.

In the limit $N \rightarrow \infty$, the spectral density is obtained from
\begin{equation}
  \rho(\lambda) =  \frac{1}{\pi} \lim_{\eta \rightarrow 0^{+}} \int_{{\rm Im} g \geq 0} d^2g \mathcal{P}(g) {\rm Im} g ,
  \label{speca}
\end{equation}  
where $d^2g \equiv d {\rm Re} g \, d {\rm Im} g $, and $\mathcal{P}(g)$ is the joint distribution of the real and imaginary parts
of $G_{ii}(z)$. The symbol $\int_{{\rm Im} g \geq 0}$ represents an integral
over $g \in \mathbb{C}$ restricted to the upper half complex plane.
The distribution of the resolvent follows from the solution of the coupled equations \cite{Rogers2008,Kuhn2008}
\begin{eqnarray}
  \mathcal{P}(g) &=& \sum_{k=0}^{\infty} p_k 
  \int_{{\rm Im} g \geq 0}  \left[ \prod_{\ell=1}^k  d^2 g_\ell \mathcal{P}_{{\rm cav}} (g_{\ell})  \right] \nonumber \\
  &\times&  \left\langle \delta \left[g -  \left(\frac{1}{z - \frac{1}{c} \sum_{\ell=1}^k J_\ell^2 g_{\ell}  } \right)\right] \right\rangle_{ J_{1},\dots,J_{k} }  , \label{res1} \\
  \mathcal{P}_{\rm cav} (g) &=& \sum_{k=0}^{\infty} \frac{k p_k}{c} \int_{{\rm Im} g \geq 0}   \left[ \prod_{\ell=1}^{k-1}  d^2 g_\ell \mathcal{P}_{{\rm cav}} (g_{\ell})  \right]  \nonumber \\
  &\times&  \left\langle  \delta \left[g -  \left(\frac{1}{z - \frac{1}{c} \sum_{\ell=1}^{k-1} J_\ell^2 g_{\ell}  } \right)\right] \right\rangle_{J_{1},\dots,J_{k-1}  }    , \label{res2}
\end{eqnarray}  
where $\langle \dots \rangle_{J_{1},\dots,J_{L}}$ denotes the average over the independent interaction strengths $J_1,\dots,J_L$, distributed according to $P_J$. The
quantity $\mathcal{P}_{\rm cav}(g)$ is the joint distribution of the real and imaginary parts of the resolvent elements $G_{ii}^{(j)}(z)$
on the cavity graph  \cite{Rogers2008,Metz2010}, defined as the graph where an arbitrary node $j$ and all its edges have been
removed. Equations (\ref{res1}) and (\ref{res2}) have been derived using both the cavity \cite{Rogers2008,Metz2010}
and the replica method \cite{Kuhn2008} of disordered systems, based on the locally tree-like structure of the configuration model.
Equations (\ref{speca}-\ref{res2}) have been rigorously proven in \cite{Bordenave2010} and they are exact in
the limit $N \rightarrow \infty$, constituting an interesting  point of departure to study the dense limit $c \rightarrow \infty$
of the spectral density $\rho(\lambda)$. 

Let $F(G)$ be an arbitrary function of the complex random variable $G$ distributed as $\mathcal{P}_{\rm cav}(g)$. By defining the average of $F(G)$
\begin{equation}
\langle F(G) \rangle = \int_{{\rm Im} g \geq 0}  d^2 g \mathcal{P}_{\rm cav}(g) F(g), \nonumber 
\end{equation}  
we can use Eq.~(\ref{res2}) and rewrite
the above expression as 
\begin{equation}
  \langle F(G) \rangle =  \int_{{\rm Im} q \geq 0} d^2 q      \mathcal{W}_{\rm cav}(q) F\left( \frac{1}{z-q} \right),
  \label{cavm} 
\end{equation}  
where we introduced the distribution of  $Q^{\prime} \equiv \frac{1}{c} \sum_{\ell=1}^{K-1} J_\ell^2 G_{\ell}$
\begin{eqnarray}
\mathcal{W}_{\rm cav} (q) &=&  \sum_{k=0}^{\infty} \frac{k p_k}{c} \int_{{\rm Im} g \geq 0}
\left[ \prod_{\ell=1}^{k-1}  d^2 g_\ell \mathcal{P}_{\rm cav} (g_{\ell})  \right] \nonumber \\
&\times& \left\langle \delta\left(q - \frac{1}{c} \sum_{\ell=1}^{k-1} J_\ell^2 g_{\ell}   \right) \right\rangle_{J_{1},\dots,J_{k-1}  }    ,
\label{ppq1}
\end{eqnarray}  
with the random variable $K$ denoting the degree of an uniformly chosen node. Following an analogous procedure, we can also rewrite  the spectral density in terms of
the distribution $\mathcal{W} (q)$ of the random variable $Q \equiv \frac{1}{c} \sum_{\ell=1}^{K} J_\ell^2 G_{\ell}$
\begin{equation}
  \rho(\lambda) = \frac{1}{\pi} \lim_{\eta \rightarrow 0^{+}}   \int_{{\rm Im} q \geq 0} d^2 q  \mathcal{W} (q) {\rm Im} \left( \frac{1}{z - q} \right),
  \label{specA}
\end{equation}  
defined as
\begin{eqnarray}
\mathcal{W} (q) &=&  \sum_{k=0}^{\infty} p_k \int_{{\rm Im} g \geq 0}
\left[ \prod_{\ell=1}^k d^2 g_\ell \mathcal{P}_{\rm cav} (g_{\ell})  \right] \nonumber \\
&\times& \left\langle \delta\left(q - \frac{1}{c} \sum_{\ell=1}^k J_\ell^2 g_{\ell}   \right) \right\rangle_{J_1,\dots,J_k} . \label{ppq2} 
\end{eqnarray}

We note that $\mathcal{W}(q)$ and $\mathcal{W}_{\rm cav}(q)$ are distributions of sums of independent complex random variables containing
a random and large number of terms. For the examples of $p_k$  in table \ref{univ}, one
  can show that, in the limit $c \rightarrow \infty$, the number of summands
  in $Q$ and $Q^{\prime}$ diverges. Thus, instead of working with the distributions of the resolvent, it is more convenient to extract the $c \rightarrow \infty$ limit of $\mathcal{W}(q)$
and $\mathcal{W}_{\rm cav}(q)$. Let us introduce  the characteristic functions $\varphi(p,t)$ and $\varphi_{\rm cav}(p,t)$ of, respectively,  $\mathcal{W}(q)$ and $\mathcal{W}_{\rm cav} (q)$
\begin{equation}
\varphi(p,t) = \sum_{k=0}^\infty  p_k \exp{\left[k S_c(p,t) \right]}, \label{car1}
\end{equation}  
\begin{equation}
\varphi_{\rm cav}(p,t) = \sum_{k=0}^\infty \frac{k p_k}{c} \exp{\left[(k-1) S_c(p,t) \right]}, \label{car2}  
\end{equation} 
with
\begin{eqnarray}
 & S_c(p,t) = \ln \Big[ \int_{{\rm Im} g \geq 0} d^2 g  \mathcal{P}_{\rm cav} (g) \int_{-\infty}^{\infty} d x P_J(x) \nonumber \\
  &\times     \exp{\left(- \frac{i p x^2 {\rm Re} g}{c} - \frac{i t x^2 {\rm Im} g}{c}  \right)} \Big].
\end{eqnarray}  
In order to study the dense limit $c\rightarrow \infty$, we expand $S_c(p,t)$ in powers of $1/c$, keeping in mind that the moments
of $\mathcal{P}_{\rm cav} (g)$ depend on $c$.
The leading term
\begin{equation}
  S_{\infty}(p,t) = -\frac{i \langle J^2 \rangle_J }{c} \left( p  \, {\rm Re} \langle G \rangle_{\infty} + t \, {\rm Im} \langle G  \rangle_{\infty} \right)
  \label{kop}
\end{equation}  
should yield the dense limit $c \rightarrow \infty$ of the spectral density. Note that we assumed
$\langle G \rangle$ converges to a well-defined limit $\langle G \rangle_{\infty}$ when $c \rightarrow \infty$.
In order to proceed further, we specify the degree distribution $p_k$ in Eqs.~(\ref{car1}) and (\ref{car2}).

\subsection{Regular and Poisson random graphs}

Although it is well-known that the dense limit of $\rho(\lambda)$ converges to the semicircle distribution for regular and Poisson
random graphs, it is instructive to illustrate our approach for these simple models, characterized by highly peaked degree
distributions around the mean value $c$.
Substituting the explicit forms of $p_k$ (see table \ref{univ}) in Eqs.~(\ref{car1}) and (\ref{car2}), and using the asymptotic
behaviour of Eq.~(\ref{kop}), we obtain
\begin{equation}
  \varphi(p,t) = \varphi_{\rm cav}(p,t) =
  e^{ - i \langle J^2 \rangle_J \left( p  \, {\rm Re} \langle G \rangle_{\infty} + t \,  {\rm Im} \langle G  \rangle_{\infty} \right)  } ,
  \label{hapq}
\end{equation}  
which promptly leads to the Dirac delta distribution in the complex plane
\begin{equation}
\mathcal{W}(q) = \mathcal{W}_{\rm cav}(q) = \delta\left( q - \langle J^2 \rangle_J  \langle G  \rangle_{\infty}  \right).
\end{equation}  
The fact that the resolvent statistics of regular and Poisson random graphs are both described by the
same characteristic function, Eq.~(\ref{hapq}), already demonstrates that these models exhibit the same
universal behaviour for $c \rightarrow \infty$.
The analytic  expression for  $\mathcal{W}_{\rm cav}(q)$ allows to determine $\langle G  \rangle_{\infty}$ through Eq.~(\ref{cavm}), which fulfils
\begin{equation}
  \langle G  \rangle_{\infty}  = \frac{1}{z - \langle J^2 \rangle_J  \langle G  \rangle_{\infty}},
  \label{hjj}
\end{equation}  
while the spectral density simply follows from Eq.~(\ref{specA})
\begin{equation}
  \rho(\lambda) = \frac{1}{\pi} \lim_{\eta \rightarrow 0^{+}}   {\rm Im} \left( \frac{1}{z - \langle J^2 \rangle_J  \langle G  \rangle_{\infty} } \right). \nonumber
\end{equation}  
By solving the quadratic equation (\ref{hjj}), we recover the Wigner semicircle law for the Gaussian ensembles of random matrix theory \cite{VivoBook}
\begin{equation}
    \rho_{\rm w} (\lambda) = 
\begin{cases}
    \frac{1}{2 \pi \langle J^2 \rangle_J } \sqrt{4 \langle J^2 \rangle_J  - \lambda^2  } ,& \text{if } |\lambda| <   2 \sqrt{\langle J^2 \rangle_J  } \\
    0 ,              & \text{if } |\lambda| \geq   2 \sqrt{\langle J^2 \rangle_J  } . \label{koaa}
\end{cases}
\end{equation}

\begin{figure}[H]
  \begin{center}
    \includegraphics[scale=0.8]{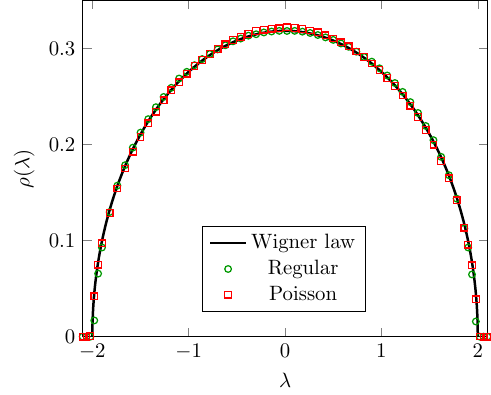}
    \caption{The dense limit $c \rightarrow \infty$ of the spectral density of regular and Poisson random networks with coupling strengths
      drawn from $P_J(x) = \delta(x-1)$. The solid line represents
      the Wigner semicircle distribution, Eq.~(\ref{koaa}), and the symbols are numerical diagonalization results obtained from $100$ samples
      of $10^4 \times 10^4$ adjacency random matrices with $c=100$.
}
\label{wigner}
\end{center}
\end{figure}

Figure \ref{wigner} compares Eq.~(\ref{koaa}) with numerical results obtained from diagonalizing large adjacency matrices $\boldsymbol{A}$ with
average degree $c=100$. The correspondence between the numerical data and the theoretical expression is excellent
for both regular and Poisson random graphs.
  
\subsection{Exponential random graphs} \label{kopiua}

In this subsection we consider the dense limit of exponential random graphs, for which $p_k$ decays slower than Poisson graphs for $k \gg 1$  (see table \ref{univ}).
Exponential degree distributions have been observed in some examples of empirical networks, such as the North American power grid \cite{Albert2004} and
the neural network of the worm {\it C. Elegans} \cite{Amaral2000}.

Inserting the expressions for $S_{\infty}(p,t)$
and $p_k$ in Eqs.~(\ref{car1}-\ref{car2}) and taking the limit $c \rightarrow \infty$, we obtain
the asymptotic forms 
\begin{align}
  \varphi(p,t) &= \frac{1}{1 + i p \langle J^2 \rangle_J {\rm Re} \langle G \rangle_{\infty} + i t \langle J^2 \rangle_J {\rm Im} \langle G \rangle_{\infty}   }, \nonumber \\
  \varphi_{\rm cav}(p,t) &= \frac{1}{\left[ 1 + i p \langle J^2 \rangle_J {\rm Re} \langle G \rangle_{\infty} + i t \langle J^2 \rangle_J {\rm Im} \langle G \rangle_{\infty}  \right]^2 }, \nonumber
\end{align}  
which already show that the dense limit of $\rho(\lambda)$ is not described by the Wigner semicircle law in this case.
The distributions $\mathcal{W}(q)$ and $\mathcal{W}_{\rm cav}(q)$ are the Fourier transforms of their  characteristic functions. Defining
\begin{align}
  \mathcal{K}(q,\epsilon) &= \int_{-\infty}^{\infty} \frac{d p \, d t}{4 \pi^2} \exp{\left( i p \, {\rm Re} q + i t \, {\rm Im} q  \right)} \nonumber \\
  &\times
  \left[ \epsilon + i p \langle J^2 \rangle_J {\rm Re} \langle G \rangle_{\infty} + i t \langle J^2 \rangle_J {\rm Im} \langle G \rangle_{\infty} \right]^{-1},
  \label{keps}
\end{align}  
with $\epsilon \geq 1$ an auxiliary parameter, $\mathcal{W}(q)$ and $\mathcal{W}_{\rm cav}(q)$ are obtained from
\begin{equation}
\mathcal{W}(q) = \mathcal{K}(q,\epsilon=1), \qquad \mathcal{W}_{\rm cav}(q) = - \frac{\partial \mathcal{K}(q, \epsilon) }{\partial \epsilon}\Bigg{|}_{\epsilon=1}.
\end{equation}  
The solution of the integral in Eq.~(\ref{keps}) is given by
\begin{align}
  \mathcal{K}(q,\epsilon) &= \frac{\Theta\left( {\rm Im} q \right)}{  \langle J^2 \rangle_J {\rm Im} \langle G \rangle_{\infty}    }
  \delta\left({\rm Re} q - {\rm Im} q  \frac{  {\rm Re} \langle G \rangle_{\infty} } {{\rm Im} \langle G \rangle_{\infty} }      \right) \nonumber \\
  &\times
\exp{\left(- \frac{\epsilon \, {\rm Im} q  }{\langle J^2 \rangle_J {\rm Im} \langle G \rangle_{\infty} }   \right)}  \nonumber ,
\end{align}   
leading to the analytic expressions
\begin{align}
   \mathcal{W}(q) &=
\frac{\Theta\left( {\rm Im} q \right)}{  \langle J^2 \rangle_J {\rm Im} \langle G \rangle_{\infty}    }
  \delta\left({\rm Re} q - {\rm Im} q  \frac{  {\rm Re} \langle G \rangle_{\infty} } {{\rm Im} \langle G \rangle_{\infty} }      \right) \nonumber \\
  &\times
\exp{\left(- \frac{{\rm Im} q  }{\langle J^2 \rangle_J {\rm Im} \langle G \rangle_{\infty} }   \right)},  \label{bb1} \\
  \mathcal{W}_{\rm cav}(q) &=
\frac{\Theta\left( {\rm Im} q \right)  {\rm Im} q  }{  \left[ \langle J^2 \rangle_J {\rm Im} \langle G \rangle_{\infty}  \right]^2  }
  \delta\left({\rm Re} q - {\rm Im} q  \frac{  {\rm Re} \langle G \rangle_{\infty} } {{\rm Im} \langle G \rangle_{\infty} }      \right) \nonumber \\
  &\times
\exp{\left(- \frac{{\rm Im} q  }{\langle J^2 \rangle_J {\rm Im} \langle G \rangle_{\infty} }   \right)},  \label{bb2}
\end{align}  
where $\Theta(\dots)$ is the Heaviside step function.

\begin{figure}[H]
  \begin{center}
    \includegraphics[scale=1.0]{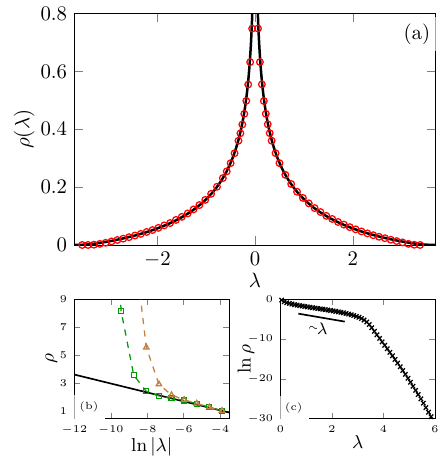}
    \caption{Dense limit of the spectral density $\rho(\lambda)$ of random networks with an exponential degree distribution and
      coupling strengths  drawn from $P_J(x) = \delta(x-1)$. (a) Comparison between the
      theoretical results (black solid lines), obtained from the numerical solutions of Eqs.~(\ref{jk1}) and (\ref{jk2}) for $\eta=0$, and numerical diagonalizations (red circles)
      of $100$ samples of  $10^4 \times 10^4$ adjacency matrices with $c=100$. (b) The logarithmic divergence of the spectral density for $|\lambda| \rightarrow 0$. The black
      solid line is the analytic result of Eq.~(\ref{gga1}), while the symbols are data derived from the numerical solutions of Eqs.~(\ref{res1}) and (\ref{res2}) for $c=80$ and
      two values of $\eta$: $\eta= 10^{-4}$ (brown triangles) and $\eta= 10^{-5}$ (green squares). (c) The exponential behaviour of the spectral density for large $|\lambda|$, obtained from the numerical solutions of Eqs.~(\ref{jk1}) and (\ref{jk2}) for $\eta=0$
}
\label{exp}
\end{center}
\end{figure}

As in the previous subsection, Eqs.~(\ref{bb1}) and (\ref{bb2}) depend  on $\langle G \rangle_{\infty}$, but this quantity can be determined
through Eq.~(\ref{cavm}). Substituting $\mathcal{W}_{\rm cav}(q)$ in  (\ref{cavm}) and
calculating the integral, we get
\begin{align}
  \langle G \rangle_{\infty} &= \frac{z}{ \langle J^2 \rangle_{J}^2 \langle G \rangle_{\infty}^2 }
  \exp{\left( - \frac{ z}{ \langle J^2 \rangle_J \langle G \rangle_{\infty}  }  \right)  } \nonumber \\
  & \times
      {\rm Ei} \left(  \frac{ z}{ \langle J^2 \rangle_J \langle G \rangle_{\infty}  }  \right)  - \frac{1}{\langle J^2 \rangle_J \langle G \rangle_{\infty}   },
      \label{resolve}
\end{align}  
with ${\rm Ei} (\dots)$ denoting the complex exponential integral function \cite{grad2007}. The solutions of the above equation yield
the dense limit of the averaged resolvent $\langle G \rangle_{\infty}$  on the cavity graph \cite{Rogers2008,Metz2010}. The expression for $\rho(\lambda)$ in terms
of $\langle G \rangle_{\infty}$ follows in an analogous way, i.e.,  
one inserts Eq.~(\ref{bb1}) in Eq.~(\ref{specA}) and calculates the remainder integral
\begin{align}
  \rho(\lambda) &= \frac{1}{\pi} \lim_{\eta \rightarrow 0^+}  {\rm Im} \Bigg[ \frac{1}{\langle J^2 \rangle_J \langle G  \rangle_{\infty}  }
    \exp{\left(- \frac{z}{ \langle J^2 \rangle_J \langle G  \rangle_{\infty} }   \right)} \nonumber \\
    &\times
    {\rm Ei} \left(   \frac{z}{ \langle J^2 \rangle_J \langle G  \rangle_{\infty} }  \right) \Bigg] ,   
\end{align}  
with $z=\lambda - i \eta$. It is suitable to introduce the dimensionless variable $\gamma(z) \equiv  \frac{z}{ \langle J^2 \rangle_J \langle G  \rangle_{\infty} } $, in
terms of which the spectral density is given by
\begin{equation}
  \rho(\lambda) = \frac{1}{\pi}  \lim_{\eta \rightarrow 0^+} {\rm Im} \left( \frac{ z^2 +  \gamma^2 \langle J^2 \rangle_J}{z \, \gamma^2   \langle J^2 \rangle_J }  \right),
  \label{jk1}
\end{equation}  
where $\gamma(z)$ solves the following  equation
\begin{equation}
  \langle J^2 \rangle_J \gamma = \frac{z^2}{\gamma^2 e^{-\gamma} {\rm Ei}(\gamma) - \gamma  }.
  \label{jk2}
\end{equation}  
Equations (\ref{jk1}) and (\ref{jk2}) constitute one of the main results of this paper. In
contrast to the resolvent Eq.~(\ref{hjj}), whose analytic solution yields the Wigner semicircle law, the fixed-point Eq.~(\ref{jk2}) for the $c \rightarrow \infty$ limit of
exponential graphs has no analytic solution for arbitrary $z$.
The spectral density obtained from the numerical solutions of Eq.~(\ref{jk2}) is shown in figure \ref{exp}-(a), together with direct
diagonalization of large adjacency matrices of exponential random graphs with $c=100$. The agreement between these two different approaches is excellent, confirming
the exactness of Eqs.~(\ref{jk1}) and (\ref{jk2}).

Figure \ref{exp}-(a) suggests that $\rho(\lambda)$ diverges at $\lambda=0$. To study
the singular behaviour of $\rho(\lambda)$ as $|\lambda| \rightarrow 0$, one needs to understand how $\gamma (z)$ vanishes as $|z| \rightarrow 0$.
With a modest amount of foresight, we make the following {\it ansatz}   
\begin{equation}
  \gamma(z) = \beta_1 \frac{z}{\sqrt{\langle J^2 \rangle_J} } + \beta_2(z) \frac{z^2}{\langle J^2 \rangle_J }  , \quad |z| \rightarrow 0,
  \label{fa}
\end{equation}  
where the function $\beta_2(z)$ is such that $\lim_{|z| \rightarrow 0} z^2 \beta_2(z) = 0 $, and the coefficient $\beta_1$ is independent of $z$. Plugging this assumption
for $\gamma(z)$ in the right hand side of Eq.~(\ref{jk2}) and expanding the result in powers of $z$ up to $O(z^2)$, we find
that  the above {\it ansatz} is consistent with Eq.~(\ref{jk2})  if $\beta_1$ and $\beta_2(z)$ are 
\begin{equation}
\beta_1 = \pm i, \quad \beta_2(z) = - \frac{1}{2} \left[ E + \ln \left(-  \frac{\beta_1 z}{\sqrt{\langle J^2 \rangle_J} }   \right) \right],
\end{equation}  
with $E$ representing the Euler-Mascheroni constant. The last step is to
substitute Eq.~(\ref{fa}) in Eq.~(\ref{jk1}) and take the limit $\eta \rightarrow 0^+$, which leads to the logarithmic divergence
\begin{equation}
  \rho(\lambda) = - \frac{1}{\pi \sqrt{\langle J^2 \rangle_J} } \left[ E + \ln \left(\frac{|\lambda|}{\sqrt{\langle J^2 \rangle_J} }   \right)  \right] 
  \label{gga1}
\end{equation}  
for $|\lambda| \rightarrow 0$. The choice of the negative sign $\beta_1 = -i$ ensures that $\rho(\lambda)$ is non-negative. A divergence
in the spectral density of random graphs at $\lambda=0$ may appear due to different reasons. For instance, the spectrum of protein-protein
interaction networks has a singularity at $\lambda=0$ due to the existence of duplicated genes \cite{Kamp2005}. In the context of
solid state physics, the density of states of a tight-binding model for a quantum particle hoping on a random graph
displays a divergence at $\lambda=0$ due to the presence of localized eigenstates \cite{Evangelou1992}.

Figure \ref{exp}-(b) compares Eq.~(\ref{gga1}) with
the numerical solutions of the distributional Eqs.~(\ref{res1}) and (\ref{res2}) for $c=80$ using a Monte-Carlo method, also known as the population dynamics
algorithm  \cite{Rogers2008,Kuhn2008,Metz2010}. 
The numerical results lie on the top of the theoretical curve up to a certain $|\lambda_{*}|$, below which the numerical data for $\rho(\lambda)$
deviates from the logarithmic behaviour of Eq.~(\ref{gga1}). As $\eta$ decreases, $|\lambda_{*}|$ shifts towards smaller values, confirming that the discrepancy
between the numerical data and Eq.~(\ref{gga1}) is an artifact due to the finite values of $\eta$ employed in the numerical
solutions of Eqs.~(\ref{res1}) and (\ref{res2}).  

As a final important property, we inspect the behaviour of $\rho(\lambda)$ for $|\lambda| \gg 1$. As illustrated in figure \ref{exp}-(c), $\rho(\lambda)$ displays a crossover from
an exponential behaviour $\ln \rho(\lambda) \propto -\lambda$ to a faster decay for increasing $|\lambda|$, indicating that the spectral density does not have sharp spectral edges, but instead
it is supported on the entire real line. This is also consistent with a stability analysis of the fixed-point Eq.~(\ref{jk2}), which shows that the complex
solution for $\gamma(z)$ remains stable for $|\lambda| \gg 1$. We point out that a perturbative study of Eq.~(\ref{jk2}) for $|\lambda| \gg 1$ is not able to capture
the analytic form describing the tails of $\rho(\lambda)$. This should not come as a surprise, as the tails of $\rho(\lambda)$ are normally caused by rare
statistical fluctuations in the graph structure \cite{Rodgers1988,Semerjian2002}.

\subsection{Borel random graphs} 

As a final example of network model, we present results for random networks with a Borel degree distribution (see table \ref{univ}). In this
case, our approach to derive analytic expressions for $\varphi(p,t)$ and $\varphi_{\rm cav}(p,t)$  is not 
useful, because the probability generating function of the Borel degree distribution does not have a closed analytic form \cite{Finner2015}.  Thus, we
obtain results through the numerical solution of Eqs.~(\ref{res1}) and (\ref{res2})  using the
population dynamics algorithm \cite{Rogers2008,Kuhn2008,Metz2010}.
For Borel random graphs, the
ratio $\sigma^2/c^2$ diverges as $c \rightarrow \infty$, which
is precisely what renders this model interesting in comparison to Poisson
and exponential graphs.

\begin{figure}[H]
  \begin{center}
    \includegraphics[scale=0.48]{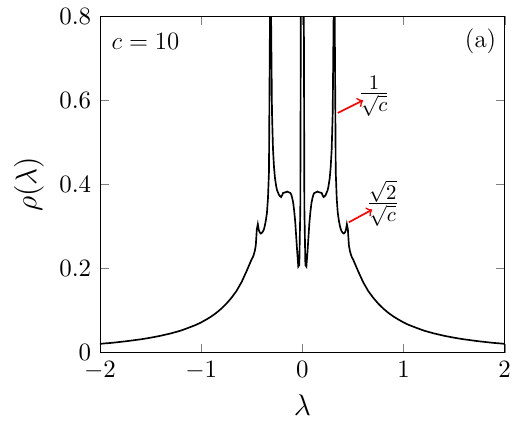}
    \includegraphics[scale=0.48]{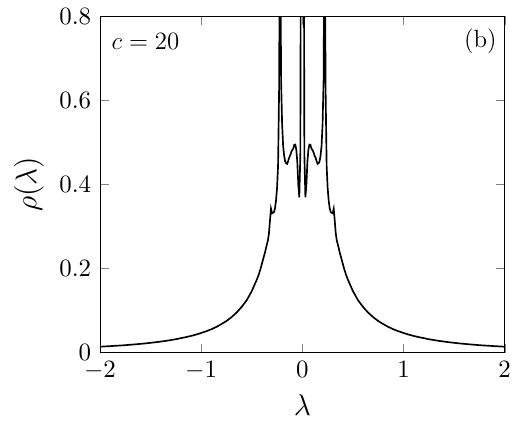}
    \includegraphics[scale=0.48]{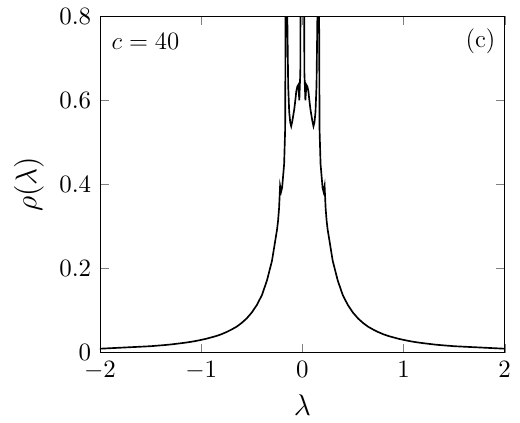}
    \includegraphics[scale=0.48]{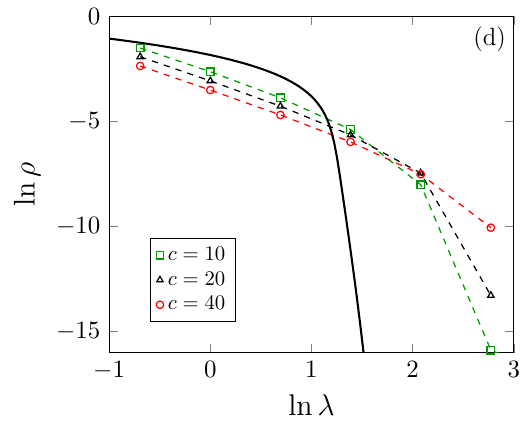}
    \caption{The spectral density of random networks from the configuration model with a Borel degree distribution (see table \ref{univ}) for different average degrees $c$ and coupling
      strengths drawn from $P_J(x) = \delta(x-1)$. All results are obtained from the numerical solutions of Eqs. (\ref{res1}) and (\ref{res2}) using the population dynamics
      algorithm with $\eta=10^{-3}$ (subfigures (a), (b), and (c)) and $\eta = 10^{-4}$ (subfigure (d)).
      Subfigures (a), (b), and (c) show the spectral density of Borel random graphs for different $c$. Figure (a) also indicates 
      the positive eigenvalues of open chains with one and two edges, and their correspondence with the position of $\delta$-peaks. 
      Figure (d) exhibits results for the tails of $\rho(\lambda)$. The symbols are results for Borel degree distributions with
      different $c$,  while the solid black curve shows the tail of $\rho(\lambda)$ for exponential dense random graphs as a comparison (see also figure \ref{exp}). The
      dashed lines are just a guide to the eye.
      }
\label{borelspec}
\end{center}
\end{figure}

Figure \ref{borelspec} illustrates the evolution of the spectral density for increasing average degree. Similarly to exponential random
graphs, the spectral density is not described by the semicircle distribution of random matrix theory, although the values of $c$ in
figure \ref{borelspec} are  not large enough to observe the dense limit of $\rho(\lambda)$.
We do not present results for larger values of $c$ than those in figure \ref{borelspec} because the population dynamics algorithm becomes
prohibitively time consuming for increasing $c$, due to the large
variance of the Borel degree distribution.

Among the distinctive features of figure \ref{borelspec}, we note the existence of $\delta$-peaks in $\rho(\lambda)$.
These peaks, often located at the eigenvalues of finite components \cite{Bauer2001,Kuhn2008,Kuhn2011,Newman2019A}, disappear
for $c \rightarrow \infty$, when the fraction of nodes belonging to the giant
component converges to one  \cite{NewmanBook}. The spectral density in figure \ref{borelspec} displays $\delta$-peaks
at the eigenvalues $\{- 1/\sqrt{c},1/\sqrt{c} \} $ and $\{ 0,-\sqrt{2}/\sqrt{c},\sqrt{2}/\sqrt{c} \}$ of the adjacency
matrices of open chains with, respectively, two and three nodes. One can show that these $\delta$-peaks  disappear in the
dense limit by computing the fraction $\Pi_s(c)$ of nodes that belongs to a finite component
with $s > 1$ nodes \footnote{See chapter 13 of \cite{NewmanBook}.}. For Poisson, exponential, and Borel degree
distributions, we obtain $\Pi_s^{({\rm pois})}(c) \propto e^{-s c}$, $\Pi_s^{({\rm exp})}(c) \propto c^{1 - 2 s}$, and $\Pi_s^{({\rm borel})}(c) \propto c^{1-s}$
for $c \gg 1$. Consequently, all network models studied in this paper do not have finite clusters in
the dense limit. In the case of the Borel degree distribution, $\rho(\lambda)$ contains
$\delta$-peaks even for very large  $c$ because the function $\Pi_s^{({\rm borel})}(c)$ decays slower for $c \gg 1$ when compared to Poisson and exponential networks.

We have also inspected the tails of the spectral density of Borel random graphs.
According to figure \ref{borelspec}-(d), the behaviour of
$\rho(\lambda)$ as a function of $\lambda$ is consistent with a power-law decay up to a certain threshold $\lambda_p$, above which the data clearly deviates from a straight line.
As a matter of fact, we have confirmed that $\rho(\lambda)$ decreases as $\eta \rightarrow 0^+$ for values of $\lambda$ larger than those appearing in figure \ref{borelspec}-(d).
In spite of that, the overall tendency of the data for increasing $c$ suggests that $\rho(\lambda)$ does decay as a power-law
in the limit $c \rightarrow \infty$.
The results in figure \ref{borelspec} are insufficient to extract the exponent characterizing the power-law
tails of $\rho(\lambda)$ for $c \rightarrow \infty$, since the spectral density has not reached its limiting behaviour for $c=40$. Taken together, these
results indicate that $\rho(\lambda)$ is supported on the entire real line.

\section{A rigorous result for the spectral density of the configuration model} \label{rigor}

In this section we state the main theorem of reference \cite{Dembo} and explain how the resolvent equations
for $c \rightarrow \infty$, obtained in the previous section, are recovered from this theorem. We use the notation of the previous sections and
we discuss the theorem in a less technical language, which
helps to establish the link between \cite{Dembo} and the results presented in this paper in a more straightforward way.

The theorem concerns the spectral density $\rho(\lambda)$ of the adjacency matrix $\boldsymbol{A}$ (see Eq.~(\ref{gaga})) of the configuration model
in the dense limit. Let $\nu(\kappa)$ be the probability density of the rescaled degrees $\frac{K_i}{c}$ ($i=1,\dots,N$)
in the limit $c \rightarrow \infty$
\begin{equation}
  \nu(\kappa) = \lim_{c \rightarrow \infty} \sum_{k=0}^\infty p_k \delta\left(\kappa - \frac{k}{c} \right).
  \label{hqmml}
\end{equation}  
The main result of \cite{Dembo} is the following theorem: \\

{\it Let $\{ K_i \}_{i=1}^N$ be i.i.d random variables forming a degree sequence of the configuration model such
  that $c=o(N)$ and $J_{ij}=1$. In the limit $N \rightarrow \infty$, if the distribution of $K_i/c$ converges to a limit distribution $\nu$ with a
  finite second moment, then  $\rho(\lambda) = \lim_{N \rightarrow \infty} \rho_N(\lambda)$
  is given by $\nu \boxtimes \rho_{\rm w}$, where
  $\rho_{\rm w}$ is the Wigner semicircle law. } \\

Thus, the spectral density $\rho(\lambda)$ of the configuration model with $c = o(N)$ is given by the free
multiplicative convolution $\boxtimes$ between the probability density $\nu$ of rescaled degrees and the Wigner
semicircle law $\rho_{\rm w}$. The free multiplicative convolution $\nu \boxtimes \rho_{\rm w}$ between the probability
measure associated to $\rho_{\rm w}$ and the probability measure corresponding to $\nu$ is the unique probability measure
associated to  $\rho$ such that its $S$-transform $S_{\rho}(\omega)$ ($\omega \in \mathbb{C}$)
fulfills the convolution property $S_{\rho}(\omega) = S_{\nu}(\omega) S_{\rm \rho_{w}}(\omega)$, where  $S_{\nu}(\omega)$
and $S_{\rm \rho_{w}}(\omega)$ are the $S$-transforms of $\nu$ and $\rho_{\rm w}$, respectively. 

The free multiplicative convolution and the $S$-transform were introduced by Voiculescu to deal
with the multiplication of free non-commuting random variables \cite{Voiculescu}. The $S$-transform finds  applications
in random matrix theory, since it is an important analytic tool to determine the spectral properties of products
of large and independent random matrices from the spectra of the individual matrices \cite{Burda2011,Nowak2015}. 

For unbounded measures on $\mathbb{R}_{+}$  and for measures with zero mean and all
moments finite, such as the Wigner semicircle distribution, the $S$-transform $S_{\varphi} (z)$ ($z \in \mathbb{C}$) of the corresponding probability density $\varphi$ is
obtained from \cite{Rao2007,Bercovici93}
\begin{equation}
  S_{\varphi}(z) = \frac{\left( 1 + z  \right)}{z} \omega_{\varphi}^{-1}(z),
  \label{jhaa}
\end{equation}  
where 
\begin{equation}
  \omega_{\varphi}(z) =  \frac{1}{z} \mathcal{C}_{\varphi}(1/z) -1.
  \label{jhii}
\end{equation}  
The function $\omega_{\varphi}^{-1}(z)$ is the functional inverse of $\omega_{\varphi}(z)$ with respect to composition and $\mathcal{C}_{\varphi}  (z)$
is the Cauchy transform of $\varphi$
\begin{equation}
\mathcal{C}_{\varphi}(z) = \int_{-\infty}^{\infty} \frac{d \lambda \varphi(\lambda)}{z-\lambda}, 
\end{equation}  
where $z$ lies outside the support of $\varphi(\lambda)$.
Instead of working directly with the inverse $\omega_{\varphi}^{-1}(z)$ in Eq.~(\ref{jhaa}), it is more convenient
to express the $S$-transform in terms of the Cauchy transform \cite{Rao2008,Nowak2015}. Since $\omega_{\varphi}^{-1} \left( \omega_{\varphi}(z)  \right) = z$, we
can rewrite Eq.~(\ref{jhaa}) as follows
\begin{equation}
  S_{\varphi} \left[ \omega_{\varphi}(z) \right] =  \frac{\left[ 1 + \omega_{\varphi}(z)  \right]}{\omega_{\varphi}(z)} z.
  \label{jupor}
\end{equation}
Hence, given the $S$-transform of a probability density $\varphi$, we are
able to uniquely determine $\varphi$.

In the context of random matrix theory, $\mathcal{C}_{\rho}(z)$ is the trace of the resolvent matrix 
\begin{equation}
  \mathcal{C}_{\rho}(z) = \lim_{N \rightarrow \infty} \frac{1}{N} \sum_{i=1}^{N} G_{ii}(z) ,
  \label{hoputa}
\end{equation}  
which, after setting $z=\lambda - i \eta$, gives access to the spectral density $\rho$
through Eq.~(\ref{popo1}). Combining Eqs.~(\ref{popo1}), (\ref{jhii}), and (\ref{hoputa}), we can also express the spectral density in terms of $\omega_{\rho}(z)$
\begin{equation}
  \rho(\lambda) = \frac{1}{\pi} \lim_{\eta \rightarrow 0^{+}} {\rm Im} \left(\frac{\omega_{\rho}(1/z) + 1 }{z}\right).
  \label{guguter}
\end{equation}  

The above theorem holds for networks where $c$ scales slowly with $N$ such that $\lim_{N \rightarrow \infty} \frac{c}{N} = 0$.
In the previous section, we have derived results by performing the limit $c \rightarrow \infty$ after
$N \rightarrow \infty$, which also implies that $\lim_{N \rightarrow \infty} \frac{c}{N} = 0$.
Consequently, with the exception of dense Borel networks for which $\nu(\kappa)$
has a divergent second moment (see table \ref{univ}),  the results for $\rho(\lambda)$ derived in the previous section should be recovered from the
above theorem.

\subsection{Regular and Poisson random graphs}

Here we apply the theorem to degree distributions $p_k$ such that $\nu(\kappa)=\delta(\kappa-1)$, like
regular and Poisson degree distributions. The procedure to derive the spectral density
from the theorem comprises three steps: first, we compute the $S$-transforms $S_{\nu}$ and $S_{\rho_{\rm w}}$; second, we
obtain the $S$-transform $S_{\rho}$ of $\rho(\lambda)$ from the convolution $S_{\rho}(\omega) = S_{\nu}(\omega) S_{\rho_{\rm w}} (\omega)$; third, we derive 
the resolvent equation from $S_{\rho}$.

Let us compute the $S$-transform of the Wigner semicircle law $\rho_{\rm w}(\lambda)$ of Eq.~(\ref{koaa}) with $\langle J^2 \rangle_{J}=1$. The
Cauchy transform $\mathcal{C}_{\rho_{\rm w}}(z)$ of $\rho_{\rm w}(\lambda)$ solves the algebraic equation \cite{Rao2008}
\begin{equation}
  \mathcal{C}_{\rho_{\rm w}}(z) = \frac{1}{z - \mathcal{C}_{\rho_{\rm w}}(z)  }.
  \label{greno}
\end{equation}  
Using Eq.~(\ref{jhii}) and rewriting Eq.~(\ref{greno}) in terms of $w_{\rho_{\rm w}}(z)$, we get
\begin{equation}
  z = \pm \frac{\sqrt{\omega_{\rho_{\rm w}}(z) }}{1 + \omega_{\rho_{\rm w}}(z) }.
  \label{huty}
\end{equation}  
Inserting the above expression in Eq. (\ref{jupor}) leads to
\begin{equation}
  S_{\rho_{\rm w}} \left( \omega_{\rho_{\rm w}} \right) = \frac{1}{\sqrt{\omega_{\rho_{\rm w}}}},
  \label{wigwig}
\end{equation}  
where we have chosen the positive sign in Eq.~(\ref{huty}).
Equation (\ref{greno}) for the Cauchy transform of $\rho_{\rm w}$ is obtained from the $S$-transform
regardless of the choice of sign in Eq.~(\ref{huty}). 

Now we turn our attention to the $S$-transform of $\nu(\kappa)=\delta(\kappa-1)$. The Cauchy transform $\mathcal{C}_\nu(z)$
of $\nu(\kappa)$ reads
\begin{equation}
  \mathcal{C}_\nu(z) = \frac{1}{z-1}.
  \label{juter}
\end{equation}
Combining Eqs. (\ref{jhii}) and (\ref{juter}), we get
\begin{equation}
z = \frac{\omega_{\nu}(z)}{1 + \omega_{\nu}(z)},
\end{equation}  
which leads to
\begin{equation}
S_{\nu}(\omega)=1
\end{equation}
after substitution in Eq.~(\ref{jupor}). According to the above theorem, the $S$-transform of the spectral
density reads
\begin{equation}
S_{\rho}(\omega) = S_{\nu}(\omega) S_{\rho_{\rm w}}(\omega) = \frac{1}{\sqrt{\omega}}, 
\end{equation}  
which immediately implies that $\rho(\lambda)$ is given by the semicircle distribution, Eq.~(\ref{koaa}).
We conclude that the spectral density of dense random networks for which the distribution of
the rescaled degrees $K_i/c$ converges to a $\delta$-peak is given by Eq.~(\ref{koaa}).

\subsection{Exponential random graphs}

For random graphs with an exponential degree distribution,  $\nu(\kappa)$
reads
\begin{equation}
    \nu(\kappa) = 
\begin{cases}
   e^{-\kappa} ,& \text{if } \kappa \geq 0,  \\
    0 ,              & \text{if } \kappa < 0 . \label{koaa1}
\end{cases}
\end{equation}
The transform $S_{\rho_{\rm w}}(\omega)$ is given by Eq.~(\ref{wigwig}), hence we just need to obtain $S_{\nu}(\omega)$. We combine the Cauchy
transform of Eq.~(\ref{koaa1}) 
\begin{equation}
\mathcal{C}_{\nu}(z) = \int_{0}^{\infty} d \kappa \frac{e^{-\kappa}}{\left( z- \kappa \right)} = e^{-z}  {\rm Ei} (z)
\end{equation}  
with Eq.~(\ref{jhii}), obtaining
\begin{equation}
  \omega_{\nu}(z) = \frac{e^{-1/z}}{z} {\rm Ei} (1/z) - 1.
  \label{uuto}
\end{equation}  
In order to derive an explicit expression for $S_{\nu}(\omega_{\nu})$ from Eq.~(\ref{jupor}), we have to invert analytically the
above equation and find $z=\phi(\omega_{\nu})$. The analytic form of $\phi(\omega_{\nu})$ is usually obtained  when the Cauchy
transform has a polynomial form \cite{Rao2008}, as in the case of the Wigner semicircle distribution. Unfortunately, here this is not the case, but we can still obtain from Eq.~(\ref{uuto}) a
self-consistent equation for the inverse function $\phi(\omega_{\nu})$ 
\begin{equation}
  \phi \left( \omega_{\nu} \right) = \frac{1}{\omega_{\nu}} \exp{\left[ -\frac{1}{\phi ( \omega_{\nu}  )  } \right] }
      {\rm Ei} \left[  \frac{1}{\phi \left( \omega_{\nu}  \right) }  \right] - \frac{\phi \left( \omega_{\nu} \right) }{\omega_{\nu}},
  \label{hopi}
\end{equation}  
while the $S$-transform follows from Eq.~(\ref{jupor})
\begin{equation}
  S_{\nu}(\omega_{\nu}) = \frac{\left( 1 + \omega_{\nu}  \right)}{\omega_{\nu}} \phi (\omega_{\nu}),
  \label{piece}
\end{equation}
with $\omega_{\nu} = \omega_{\nu}(z)$. The solution of Eq.~(\ref{hopi}) determines $S_{\nu}(\omega_{\nu})$ through Eq.~(\ref{piece}).

Now we are ready to recover the equations determining $\rho(\lambda)$ of exponential random graphs  using the
theorem of \cite{Dembo}. The $S$-transform of $\rho(\lambda)$ follows from the convolution
\begin{equation}
S_{\rho}(\omega_{\rho}) = S_{\nu}(\omega_{\rho}) S_{\rho_{\rm w}}(\omega_{\rho}) = \frac{\left( 1 + \omega_{\rho}  \right)}{\omega_{\rho} \sqrt{\omega_{\rho}}} \phi(\omega_{\rho}).
\end{equation}  
From Eq.~(\ref{jupor}) we conclude that $\phi(\omega_{\rho})$ must also fulfill
\begin{equation}
z \sqrt{\omega_{\rho}}  = \phi(\omega_{\rho})
\end{equation}  
for $z$ outside the support of $\rho(\lambda)$.
Substituting the above relation  in Eq. (\ref{hopi}), we find an equation that determines $\omega_{\rho}(1/z)$
\begin{align}
  \frac{\sqrt{\omega_{\rho}(1/z)} }{z} &= \frac{1}{\omega_{\rho}(1/z) } \exp{\left[ - \frac{z}{\sqrt{\omega_{\rho}(1/z)  } }   \right]}
       {\rm Ei} \left[ \frac{z}{\sqrt{\omega_{\rho}(1/z)  } }   \right] \nonumber \\
  &- \frac{1}{z  \sqrt{\omega_{\rho}(1/z)} }. \label{kopiop}
\end{align}  
The solution for $\omega_{\rho}(1/z)$ at $z=\lambda-i \eta$ allows to compute the spectral density using Eq. (\ref{guguter}). Setting
\begin{equation}
\gamma(z) = \frac{z}{\sqrt{\omega_{\rho}(1/z)}},
\end{equation}  
we conclude that Eqs.~(\ref{guguter}) and (\ref{kopiop}) are identical to Eqs. (\ref{jk1}) and (\ref{jk2}), respectively.
We also identify $\sqrt{\omega_{\rho}(1/z)} = \langle G \rangle_{\infty}$ as the averaged resolvent on the cavity graph.

\section{The fourth moment of the eigenvalue distribution} \label{assa}

The results of sections \ref{bbba} and \ref{rigor} show that the dense limit of $\rho(\lambda)$ depends
on the degree distribution of the configuration model, and the
scope of the Wigner semicircle law is limited to graphs where $p_k$ becomes highly concentrated around the mean degree for $c \rightarrow \infty$. In this section we derive a simple
equation that reveals the influence of degree fluctuations on the spectral density, giving  an exact
condition for the breakdown of the Wigner semicircle law. We end up this section by proposing a classification
of the different universal behaviours of the dense limit of $\rho(\lambda)$ for the configuration
model of networks.

It is natural to characterize the statistics of random variables by computing
the ratio between their moments.
In the case of locally tree-like random networks, the odd moments of $\rho(\lambda)$
are zero and the simplest dimensionless parameter of this type
is the kurtosis $K_N(c)$
\begin{equation}
  K_N(c) = \frac{\int_{-\infty}^\infty d \lambda \lambda^4 \rho_N(\lambda)  }{\left[ \int_{-\infty}^\infty d \lambda \lambda^2 \rho_N(\lambda) \right]^2 } =
 \frac{ N   {\rm Tr} \boldsymbol{A}^4 }{   \left( {\rm Tr} \boldsymbol{A}^2  \right)^2 },
\end{equation}  
where we used the definition of the empirical spectral measure, Eq.~(\ref{jkla}).
When $J_{ij}= \sqrt{c} \,\, \forall i,j$, the trace ${\rm Tr} \boldsymbol{A}^n$ ($n=2,3,\dots$) is the total number of closed walks of length $n$ in the graph, where the length
of a walk between nodes $i$ and $j$ is the number of edges that a walker traverses when going from one node to the other \cite{VanBook}. The relation between
the moments of $\rho_N(\lambda)$ and ${\rm Tr} \boldsymbol{A}^n$ is very important in spectral graph theory, as it connects the eigenvalue statistics with
the network structural properties  \cite{VanBook}. The
limit $N \rightarrow \infty$ must be taken before the $c \rightarrow \infty$ limit.

The trace of the second power of the adjacency matrix $\boldsymbol{A}$ reads
\begin{equation}
{\rm Tr} \boldsymbol{A}^2  = \frac{1}{c} \sum_{i=1}^N \sum_{j \in \partial_i} J_{ij}^ 2, 
\end{equation}  
where $\partial_i$ represents the set of nodes connected to $i$. For $N \rightarrow \infty$, we obtain
\begin{equation}
  \lim_{N \rightarrow \infty} \frac{1}{N} {\rm Tr} \boldsymbol{A}^2  = \frac{1}{c} \left\langle  \sum_{j \in \partial_i} J_{ij}^ 2  \right\rangle_{J,K} = \langle J^2 \rangle_J,
  \label{jqqpo}
\end{equation}  
with  $\left\langle  \dots  \right\rangle_{J,K}$ denoting the ensemble average over the degrees and the coupling strengths. The trace
of the fourth power can be written as
\begin{align}
  c^2 {\rm Tr} \boldsymbol{A}^4 &= 2 \sum_{i=1}^{N}  \sum_{j,r \in \partial_i}   J_{ij}^2 J_{ri}^2 - \sum_{i=1}^{N}   \sum_{j \in \partial_i}  J_{ij}^4 \nonumber \\
  &+  \sum_{i=1}^{N} \sum_{j \in \partial_i} \sum_{k \in \partial_j \setminus i} \sum_{r \in \partial_k \setminus j} C_{ir} J_{ij} J_{jk} J_{kr} J_{ri}  ,
  \label{gaap6}
\end{align}  
where $\partial_j \setminus i$ is the set of nodes adjacent to $j$, except for node $i \in \partial_j$. In the limit $N \rightarrow \infty$, the configuration model
has a locally tree-like structure, cycles of length four are rare, and hence the last term on the right hand side of Eq.~(\ref{gaap6}) gives
only a subleading contribution, which can be neglected for $N \rightarrow \infty$, yielding the expression
\begin{equation}
  \lim_{N \rightarrow \infty} \frac{1}{N} {\rm Tr} \boldsymbol{A}^4
  = \frac{2}{c^2} \langle K (K-1)  \rangle_K \langle J^2 \rangle_J^2  + \frac{1}{c} \langle J^4 \rangle_J.
\end{equation}  
Thus, we obtain an exact analytic expression for the $N \rightarrow \infty$ limit of the kurtosis
\begin{equation}
  K_{\infty}(c) = \frac{1}{c} \frac{ \langle J^4 \rangle_J  }{\langle J^2 \rangle_J^2   } + \frac{2}{c} (c-1) + 2 \frac{\sigma^2}{c^2},
  \label{jjuu}
\end{equation}  
valid for arbitrary degree distributions with first and second moments finite. Note that $K_{\infty}(c)$ is invariant under a rescaling of the adjacency
matrix elements. Equation (\ref{jjuu})
shows how the kurtosis of the eigenvalue distribution is linked to the degree fluctuations.

Let us now establish some general conclusions about the dense limit of $\rho(\lambda)$. For $c \rightarrow \infty$,  $K_{\infty}(c)$ behaves as
\begin{equation}
  \lim_{c \rightarrow \infty} K_{\infty}(c) = K_{\rm w} \left( 1 +  \lim_{c \rightarrow \infty} \frac{\sigma^2}{c^2} \right),
  \label{hhaap}
\end{equation}  
where $K_{\rm w} = 2$ is the kurtosis of the Wigner semicircle distribution, Eq.~(\ref{koaa}).
The kurtosis of the distribution $\rho_{\rm w}(\lambda)$ follows from its moments, which are given by the Catalan numbers  \cite{Wigner55}.
Equation (\ref{hhaap}) unveils the central role of the degree fluctuations to the $c \rightarrow \infty$ limit of $\rho(\lambda)$.
It follows that
\begin{equation}
  \lim_{c \rightarrow \infty} \frac{\sigma^2}{c^2} > 0
  \label{hapqoe}
\end{equation}  
is a {\it sufficient} condition for the breakdown of the Wigner semicircle law. In the
previous section we have shown that $\rho(\lambda)$ converges to the semicircle distribution when the
distribution $\nu$ of rescaled degrees is a Dirac $\delta$-peak, which is fully consistent with Eq.~(\ref{hapqoe}). 
One naturally expects that the eigenvalue statistics of network models that fulfil condition (\ref{hapqoe}) is not described
by the traditional results of random matrix theory.

Based on Eq.~(\ref{hhaap}), we can also put forward a classification of the different universal behaviours of $\rho(\lambda)$
in the dense limit. Let us consider weighted undirected networks, defined by the adjacency matrix of Eq.~(\ref{gaga}), in
which the variance of $p_k$ scales as $\sigma^2 = H c^{\alpha}$ for large $c$, with $H > 0$ and $\alpha$  arbitrary parameters.
For this broad class of network models, the different universal behaviours of $\rho(\lambda)$ can be
classified in terms of the exponent $\alpha$ that quantifies the strength of the degree fluctuations.
For $\alpha < 2$, the relative variance $\sigma^2/c^2$ vanishes for $c \rightarrow \infty$, which is a strong
indication that $p_k$ is highly peaked around $c$. Thus, we obtain $\lim_{c \rightarrow \infty} K_{\infty}(c) = 2$ for $\alpha < 2$, and it is
reasonable to conjecture that the dense limit of  $\rho(\lambda)$ is given by the Wigner semicircle law.
Regular and Poisson random graphs are the main examples of this class of random networks characterized by $\alpha < 2$. For $\alpha > 2$, the
limit $\lim_{c \rightarrow \infty} K_{\infty}(c)$ diverges and we expect $\rho(\lambda)$
decays as a power-law $\rho(\lambda) \sim |\lambda|^{-\beta}$ for $|\lambda| \gg 1$, with an exponent $3 < \beta < 5$. The lower bound on $\beta$ is due
to the finite variance of $\rho(\lambda)$ in the dense limit (see Eq.~(\ref{jqqpo})). Borel random graphs are characterized by $\alpha > 2$.
Finally, network models with $\alpha=2$ have a finite kurtosis $\lim_{c \rightarrow \infty} K_{\infty}(c) = 2(1 +  H) > K_{\rm w}$, whose precise value is determined by the prefactor $H$.
Exponential random networks belong to this later class where $\alpha=2$ and $H=1$. The results presented in section \ref{bbba} are entirely consistent with this classification scheme.

\section{Final remarks} \label{gugu}

Traditional random matrix theory deals with the spectral properties of large random matrices with
independent and identically distributed elements, providing a theoretical framework to address the universal properties
of large interacting systems \cite{AkemannBook}. It is widely known that the eigenvalue distribution of undirected random networks strongly deviates from the
Wigner semicircle law of random matrix theory in the {\it sparse} regime \cite{Rodgers1988,Bauer2001,Farkas2001,Doro2003,Rogers2008,Kuhn2008,Newman2019A}, i.e., when the
average degree $c$ is finite. As $c$ grows to infinity, random networks
gradually become more fully-connected and one may expect that random matrix theory describes well the spectral
properties of  {\it dense} random networks. We have shown in this paper that this is generally not the case.

We have studied the spectral density of the adjacency matrix of networks drawn from the
configuration model  with four distinct degree distributions. Our main conclusion is that the dense limit of the
spectral density is governed by the strength of the degree fluctuations. It turns out that the semicircle distribution
of random matrix theory is recovered only when the degree distribution becomes, in the dense
limit, highly concentrated around the mean degree. We have also derived an exact relation between
the fourth moment of the eigenvalue distribution and the variance of the degree distribution, from which
a sufficient condition for the breakdown of the Wigner semicircle law follows. We point out that  the degree distributions considered
in this work have exponentially decaying tails, implying that all moments of the degree distribution
are finite (the moments diverge only in the dense limit). 

From the results derived in this work, one expects that, in general, the circular law of random matrix
theory \cite{Rogers2009,Metz2019} should also fail in describing the spectral density of directed random
networks in the dense limit.
Following the techniques discussed in sections \ref{bbba} and \ref{assa}, the study of the eigenvalue distribution of  directed networks in the dense limit is just around the corner.

Among the results in section \ref{bbba}, we highlight the analytic equation for the averaged
resolvent (see Eq.~(\ref{resolve})), which determines the spectral density
of exponential random graphs in the dense limit. This equation has no analytic solution, in
contrast to the analogous Eq.~(\ref{hjj}) yielding the Wigner semicircle distribution.
We have derived important features of the
spectral density of dense exponential graphs, such as the existence of a logarithmic singularity
around the origin and the absence of sharp spectral edges. In the case of dense random networks with a Borel degree
distribution, we have studied the spectral density by solving numerically the exact Eqs.~(\ref{res1}) and (\ref{res2}) for large
values of the average degree.
The results in figure \ref{borelspec} indicate that the spectral density
of dense Borel networks is also supported on the entire real line, exhibiting power-law tails for large eigenvalues.
Taken together, these results reveal remarkable differences with respect to the Wigner semicircle law.
 
We have shown how the analytic results of section \ref{bbba} are recovered from the main theorem
in reference \cite{Dembo}. While the results of section \ref{bbba} follow from the equations for the distribution of
the resolvent, the theorem in \cite{Dembo} is proved through a series of techniques, including
tools from free probability theory. These two approaches are fundamentally  different and we hope
that our work stimulates research towards a rigorous proof of their equivalence for arbitrary degree distributions.
We also remark that, although the theorem in \cite{Dembo} is a compelling result about the spectral density, the approach of section \ref{bbba}
is also very interesting, since it allows to compute the distribution of the resolvent, the distribution of the self-energy \cite{Abou1973}, and the inverse participation
ratio \cite{Metz2010} in the dense limit. All these quantities are important, for instance, to study the localization
properties of eigenvectors.

Based on Eq.~(\ref{hhaap}) for the fourth moment of the eigenvalue distribution,  we have put forward a classification scheme of the different
universal behaviours of the spectral density in the dense regime. The classification  holds for network
models in which the variance $\sigma^2$ of the degree distribution scales as $\sigma^2 \propto c^{\alpha}$ for $c \gg 1$.
Networks with $\alpha < 2$ exhibit weak degree fluctuations and we have conjectured that the spectral density converges to the Wigner
semicircle distribution for $c \rightarrow \infty$. Networks with $\alpha > 2$ display strong degree fluctuations and the spectral density
is characterized by a divergent fourth moment and power-law tails for $c \rightarrow \infty$. Finally, the spectral density of
dense networks with $\alpha=2$ has a finite fourth moment, whose value is larger than in the case of the Wigner semicircle
distribution. The results for the specific degree distributions in section \ref{bbba} are entirely consistent with this classification scheme.
In light of this, it would be
interesting to design a network model with a degree distribution that has finite moments and interpolates among the different classes.

Overall, our results shed light on the fundamental role of the degree fluctuations in the dense limit of random networks. In this
context, it would be interesting to inspect the role of degree fluctuations in the local spectral properties \cite{Lee2018,Huang2020,Bauer2020} of
dense random networks, since the local statistics of the spectrum usually exhibits a higher level of universality in comparison to the global statistics.
The condition for the breakdown of the Wigner semicircle law, Eq.~(\ref{hapqoe}), should also apply beyond the realm of network
spectra, indicating whether the degree fluctuations are strong enough to cause the failure
of classic mean-field models on dense networks, such as the Curie-Weiss model \cite{BaxterBook} and the Sherrington-Kirkpatrick model \cite{Sherrington1975}.
Thus, we expect our results will stimulate research towards a better understanding of the dense limit of various
models defined on random graphs, including  ferromagnetic and spin-glass models \cite{Leone2002}, social dynamics \cite{Castellano2009}, neural networks \cite{CoolenBook}, and
synchronization \cite{Restrepo2005}.

\section*{acknowledgements}

We thank an anonymous referee for pointing out reference \cite{Dembo}, which has led to section \ref{rigor}. FLM thanks London Mathematical Laboratory and
CNPq/Brazil for financial support. JDS acknowledges a fellowship from the London Mathematical Laboratory.

\bibliography{biblio.bib}

\end{document}